\documentclass[preprint,11pt]{article}

\usepackage{graphicx}
\usepackage{dcolumn}
\usepackage{bm}
\usepackage{color}

\newcommand {\lab}[1]{\label{eq:#1}}

\newcommand {\be}[1]{\begin{equation}{\lab{#1}}}
\newcommand {\ee}{\end{equation}}
\newcommand {\bea}{\begin{eqnarray}}
\newcommand {\eea}{\end{eqnarray}}

\begin{document}

\title{ Stages of dynamics in the Fermi-Pasta-Ulam 
system as probed by the first Toda integral}

\author{
\textbf{ H. Christodoulidi$^{1}$, C. Efthymiopoulos$^{1}$}\\
$^{1}$Research Center for Astronomy and Applied Mathematics  
\\ Academy of Athens, Athens, Greece}


 \maketitle


\begin{abstract}
We investigate the 
long term evolution of trajectories in the Fermi-Pasta-Ulam (FPU) system, 
using as a probe the first non--trivial integral $J$ in the hierarchy of  
integrals of the corresponding Toda lattice model. To this end we perform 
simulations of FPU--trajectories for various classes of initial conditions 
produced by the excitation of isolated modes, packets, as well as `generic' 
(random) initial data. For initial conditions corresponding to localized 
energy excitations, $J$ exhibits variations yielding `sigmoid' curves
 similar to observables used in literature, e.g. the `spectral 
entropy' or various types of `correlation functions'. However, $J(t)$ is free of 
fluctuations inherent in such observables, hence it constitutes an ideal 
observable for probing the timescales involved in the stages of FPU dynamics. 
We observe two fundamental timescales: i) the 
`time of stability' (in which, roughly, FPU trajectories behave like Toda),  
and ii) the `time to equilibrium' (beyond which energy equipartition is 
reached). Below a specific energy crossover, both times are found to scale 
exponentially as an inverse power of the specific energy. However, this 
crossover goes to zero with increasing the degrees of freedom $N$ as 
$\varepsilon _c \sim N^{-b}$, with $b \in [1.5, 2.5]$. For `generic data' 
initial conditions, instead, $J(t)$ allows to quantify the continuous 
in time slow diffusion of the FPU trajectories in a direction transverse 
to the Toda tori. 
\end{abstract}


\maketitle

\section{Introduction}
\label{intro}
The numerical experiment of Fermi, Pasta and Ulam \cite{feretal1955,
tsingou} in 1954 aimed to probe ergodicity in an one--dimensional chain of 
$N$ weakly nonlinearly coupled oscillators. The discovery of FPU recurrences 
led to a substantial open question in Statistical Mechanics: Does a system 
with $N$ degrees of freedom, stemming from a generic perturbation to an 
integrable model, always tend to an ergodic state as $N$ becomes large?  
Several numerical studies confirm ergodicity in the FPU problem for large $N$, 
and for various special types of initial conditions \cite{benettin1,Livii,kantz}. Nevertheless, 
despite a vast effort we are still far from reaching a rigorous answer to the 
above question. One obstruction to rigorous results stems from difficulties 
in implementing perturbation theory in the `thermodynamic limit', i.e. when 
the energy per oscillator $\varepsilon$ (specific energy) is small and fixed 
while $N$ becomes large.  
 
Detailed reviews on the FPU problem can be found in \cite{bergman,FPUbook}. 
We quote here some important numerical and theoretical main lines of approach 
to the FPU problem, which are related to our present study:

i) {\it Departure from the harmonic limit}: the lowest order integrable 
approximation to the FPU is the chain of uncoupled oscillators constituting 
its linear normal modes. Initial conditions `{\it {\` a} la Fermi}', i.e., 
low--frequency normal mode excitations, have been studied extensively 
\cite{Ponno2,Livi2,benettin1,Flach2005,flaetal2006,
flapon2008,gentaetal,Ponno1,dresden,Livi}. They are well known to lead to exponentially localized energy 
profiles in the space of normal modes. Such profiles persist for very long 
times (`metastable states'), but eventually evolve to states closer to energy 
equipartition (`equilibrium states'). In the harmonic limit (small specific 
energies) this behavior can be interpreted through the stability properties 
of particular low-dimensional invariant objects of the FPU phase space. 
Most notably, the `$q$--breathers' are Lyapunov periodic orbits forming the 
continuation of the linear modes \cite{Flach2005,flaetal2006,PF}. On the other hand, the `$q$--tori' 
correspond to the extension, in the nonlinear regime, of quasi-periodic 
motions pertinent to excitations of packets of normal modes in the linear 
regime \cite{tori1,tori2,tori3}. As far as the $q$--breathers or $q$--tori are 
stable (e.g. with respect to transverse perturbations), the metastability 
phenomenon can be understood as a case of `stickiness' around these objects. 
From this viewpoint, however, it remains puzzling that the metastability 
phenomenon {\it persists for specific energies higher} than the stability 
thresholds of $q$--breathers or $q$--tori \cite{Flach2005,flaetal2006,tori1,tori3}, 
and clearly beyond the harmonic regime (although of course still small, i.e. $\varepsilon <1$). 
One should note here that slow 
in time deviation from a quasi-integrable behavior seems to occur also in 
cases of more general types of initial conditions, where, from the beginning, 
the energy is distributed in the whole spectrum, instead of isolated packets 
of modes. In this case, we can compute the evolution of the 
{\it autocorrelation functions} for suitably defined phase space quantities 
\cite{carati,carati2,CMGA,carati3}. The equilibrium state is identified as the point of 
complete decay of the autocorrelation coefficients.

ii) {\it Stochasticity threshold:} In 1959 Chirikov published his celebrated 
works \cite{chirikov1,chirikov2,IC} on the existence of a stochasticity 
threshold in terms of specific energy in the FPU, which separates chaotic 
from weakly chaotic or regular motions. Chirikov's threshold $\varepsilon _{s}$ 
is derived by computing conditions under which the FPU resonances exhibit 
substantial overlapping. He finds that $\varepsilon _{s}$ vanishes with $N$ 
like $\varepsilon _{s}\sim 1/N^4$. In \cite{Shepe}, Shepelyansky extended 
Chirikov's results by including the case of low--mode  excitations. Thus, 
the general conclusion from such studies is that one should expect 
stochasticity to prevail, and metastability phenomena to disappear, in the 
thermodynamic limit. Similar conclusions are reached when considering the 
dependence of the stability thresholds on $N$ for the $q$--breathers 
\cite{Flach2005,flaetal2006} and a weaker decay of the form $1/N$ for the 
$q$--tori \cite{tori3}.

However, as noted in \cite{Shepe}, any approach based on general criteria 
of resonance overlap cannot exclude the possibility that the system under 
investigation is very close to an integrable one, in which 
case  estimates on resonances do not apply. In the words 
of Shepelyansky: `This point is very crucial for the $\alpha $--FPU problem, 
since at low energy it is very close to the Toda lattice. Due to that 
generally we should expect that, in contrast to the above estimates and 
numerical data the dynamics of the $\alpha $--FPU problem will be integrable'. 
In recent years, there has been increasing attention to Toda as a reference 
integrable model close to the FPU \cite{benettin2,benettin3,tori3,gold,dresden}. 
Already in 1982 the pioneering work of Ferguson et al. \cite{ferguson} 
put forward the idea that the higher polynomial order contact between FPU and 
Toda allows to associate the FPU's integrable--like behavior (e.g. FPU 
recurrences) with Toda. In fact, this can be regarded as a `discrete' analogue 
of the Zabusky--Kruskal approach, which associates the FPU with a different 
integrable continuous limit, i.e., the Korteweg--de Vries (KdV) equation 
\cite{kdv}.

In the present paper, we propose a simple method to measure  the proximity 
of (evolving) FPU states, generated by various types of initial conditions, 
to the dynamics of the nearby integrable Toda model: this is to observe 
directly the evolution of the functions $J(q,p)$ yielding the Toda integrals, 
for a single or an ensemble of FPU trajectories. An independent work along 
the same direction appeared recently in \cite{gold}. In the sequel, we focus 
on the evolution of only the first non-trivial Toda integral (besides the 
energy), denoted hereafter as $J$. One has a constant value $J(t)=c$ for 
any trajectory under the exact Toda dynamics, while $J(t)$ varies along 
the FPU trajectories. As shown below, these variations allow to characterize 
the FPU stages of dynamics according to their proximity to the Toda dynamics. 
In particular, one can clearly identify `stickiness' effects, and measure 
the times up to which the FPU trajectories remain sticky to nearby Toda 
tori.

Examples of this behavior, along with a corresponding quantitative analysis, 
are given in cases covering most widespread categories of initial conditions 
encountered in FPU literature. These roughly cover three classes of initial 
conditions: i) single mode or packet of modes excitations with coherent or 
random phases, ii) random data, where the whole energy spectrum has comparable 
power from the start, and iii) generic data close to energy equipartition. 
Cases (i) and (ii) are characterized by distinct phases of evolution of the 
FPU trajectories, which roughly correspond to the stages of dynamics 
recognized in \cite{Ponno2,dresden,benettin2}. We show below how $J(t)$ allows to 
measure the timescales related with the stages of dynamics as well as the 
latter's dependence on the system's specific energy and number of degrees 
of freedom. In case (iii), using $J(t)$ we provide direct evidence of 
diffusion taking place as the FPU trajectories wander in phase space 
transversally to `Toda tori'. This is useful also in 
interpreting the observed decay of the autocorrelation for the Toda 
integrals as reported in \cite{benettin2}. 

The paper is organized as follows: Section 2 deals with definitions and 
examples related to the utility of $J$ as an observable for featuring 
FPU--evolution. Section 3 analyses the information obtained by $J$ in 
numerical experiments covering each of the aforementioned class of initial 
conditions. Section 4 contains the basic conclusions of the present study.

\section{The first Toda integral as an FPU--observable}
\label{nmti}
\begin{figure}
\centering
\includegraphics[width=0.7\textwidth]{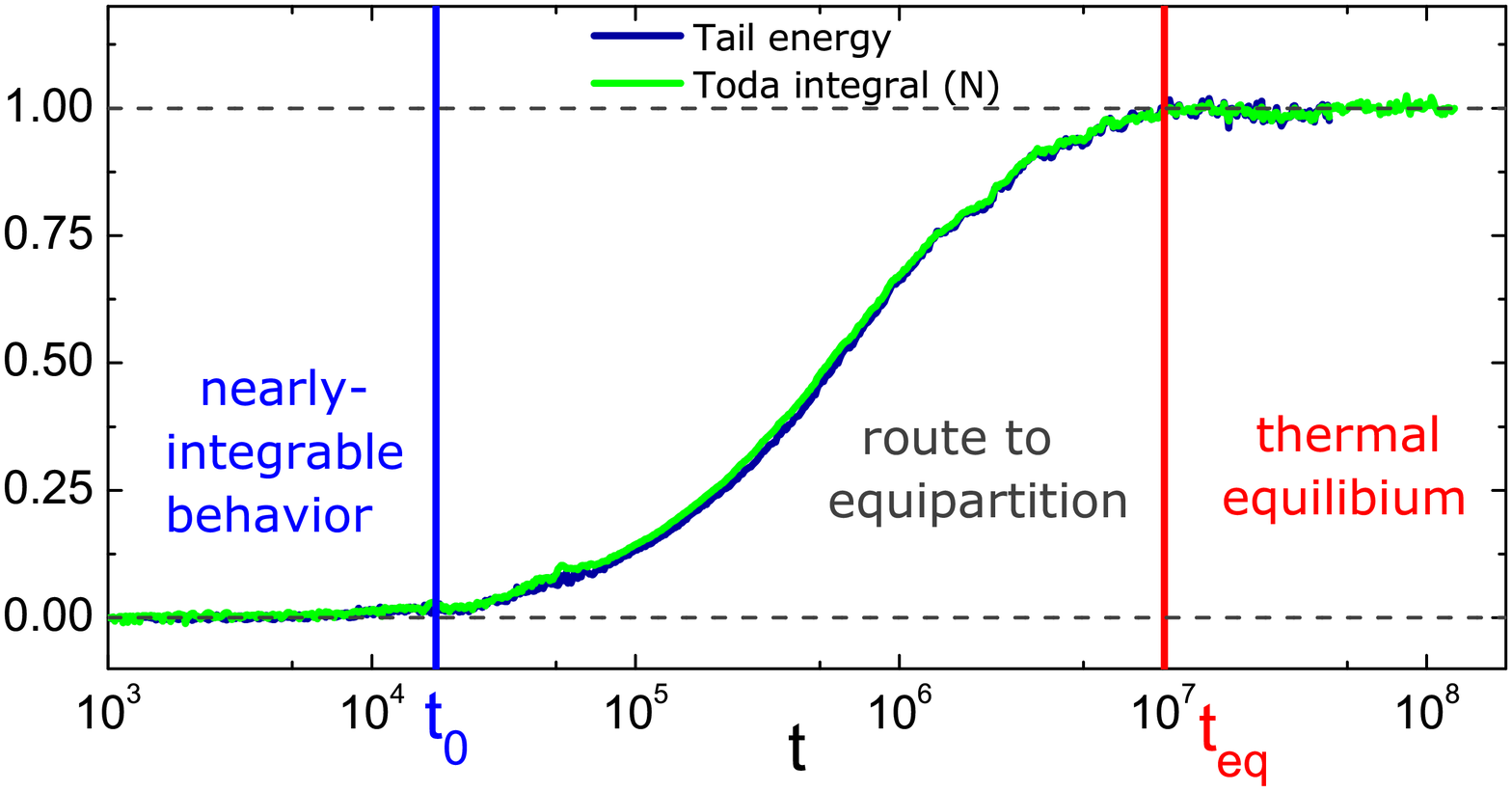}\\
\caption{Evolution of the normalized Toda integral $J$ and of the tail energy 
$\eta$ when exciting the first $ 12.5\%$ of the normal modes 
in the system with $N=8192$, $\alpha =1/2$ and $ \varepsilon =0.01$. A `nearly integrable' behavior 
(near constancy of $J(t)$) appears for $0\leq t\leq t_0$, a sigmoid increase 
of both indicators afterwards, leading to a new stabilization after $t\geq 
t_{eq}$. The times $t_0$ and $t_{eq}$ denote the `stability time' and 
`equilibrium time' respectively.}
\label{overview}
\end{figure}
\begin{figure}
\centering
\includegraphics[width=0.6\textwidth]{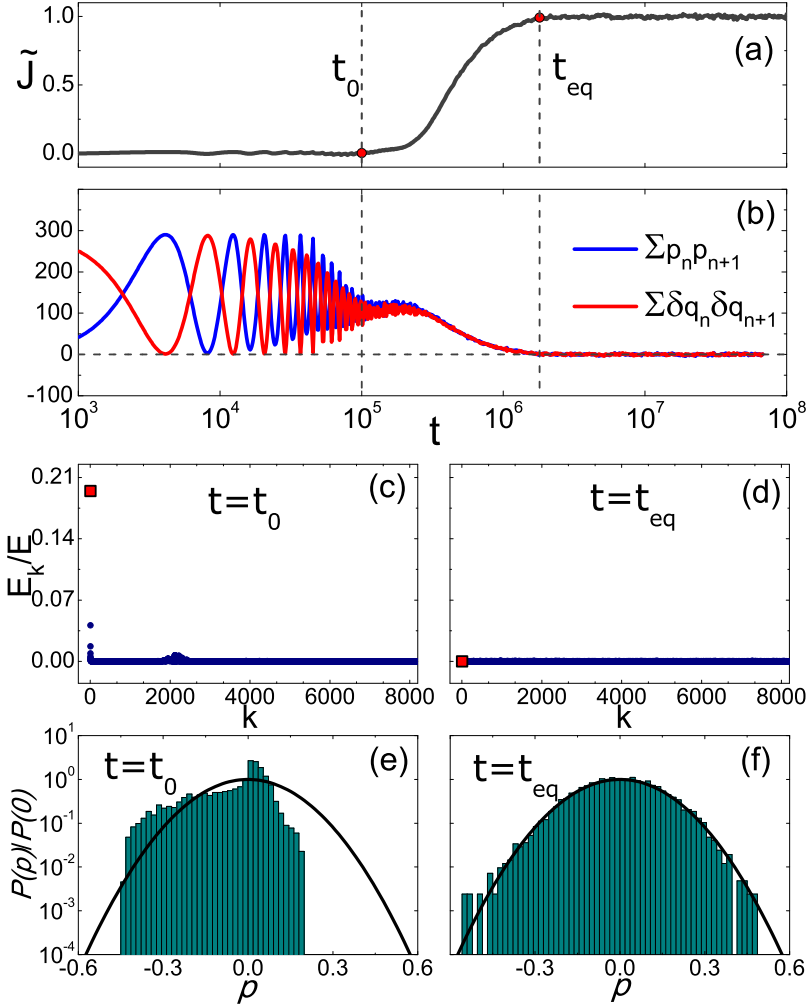}
\caption{  Excitation of the first normal mode.
(a) Evolution of the normalized Toda integral $J$. (b) Evolution of the sums 
$\sum p_np_{n+1}$, $\sum \delta q_n\delta q_{n+1}$ as indicators of 
correlation between the real space variables. (c) and (d):  The energy spectrum $E_k/E$ at  $t=t_0$ and 
$t=t_{eq}$ respectively.
(e) and (f): The momenta 
distributions at $t=t_0$ and $t=t_{eq}$ respectively. The black curve 
indicates a Gaussian distribution with dispersion $\sigma_p^2=2\varepsilon$.  }
\label{eq}
\end{figure}
 
Fermi, Pasta and Ulam \cite{feretal1955} considered the one--dimensional 
lattice consisting of $N$ weakly nonlinearly coupled oscillators. The 
dynamics of this model is described by the Hamiltonian:
\begin{eqnarray}\label{fpuham} 
H_{FPU}=\sum_{n=0}^{N-1} \big[ {1\over 2} p_n^2 
+ {1 \over 2} (q_{n+1}-q_n)^2 + {\alpha \over 3} (q_{n+1}-q_n)^3 \big]
\end{eqnarray}
where $q_n$ is the $n$--th particle's displacement with respect to 
equilibrium and $p_n$ its canonically conjugate momentum. Fixed boundary 
conditions are imposed for $q_0=q_{N}=p_0=p_N=0$. 

The normal modes of the harmonic limit $\alpha =0$ are defined by the 
canonical transformation 
$(Q_k,P_k) = 
\sqrt{2\over N}\sum_{n=1}^{N-1} (q_n,p_n) \sin\left({n k\pi\over N}\right)$.    
The normal mode energies are 
$E_k= {1\over 2} (P_k^2+ \omega_k ^2 Q_k^2)$
with frequencies $\omega _k = 2 \sin( k \pi /2N)$, $k=1,\ldots,N-1$.

The FPU Hamiltonian has a third--order polynomial contact with the 
integrable Toda Hamiltonian \cite{Toda}
\begin{eqnarray}  \label{B1a}
H_{T}=  \sum_{n=0}^{N-1} \big[ {1\over 2} p_{n}^{2} +  
\frac{ 1}{4\alpha^{2}} (a_n^2  -1)\big],
\end{eqnarray}
where $a _n=e^{\alpha (q_{n+1}-q_n)}$. Taylor--expanding the 
exponential $a_n$, one obtains $H_T=H_{FPU}+{\cal O}([\delta q_n]^4)$,
where $\delta q_n=q_{n+1}-q_n$ are the relative displacements.

The expressions of all the Toda integrals are given in \cite{Flaschka,Henon}. 
Besides $H_T$, the first non--trivial Toda integral after adding and dividing 
with some constants, takes the form:
\begin{eqnarray}\label{j1}
J={1 \over N} \sum _{n=0}^{N-1}  \big[ {\alpha ^2 \over 2} p_n^4 +  
\frac{ 1 }{2}a_n^2 (p_n^2 + p_n p_{n+1}+ p_{n+1}^2 ) 
+\frac{a_n^2}{16 \alpha ^2}  ( a_{n+1}^2+a_n^2 +a_{n-1}^2) 
- \frac{3} { 16 \alpha ^2} \big]   
\end{eqnarray}
where fixed boundary conditions are imposed for $a _{0}=a _{-1}
=e^{\alpha q_1}$, $a _{N}=a _{N-1}=e^{-\alpha q_N}$.

We outline the behavior of $J$ along FPU trajectories through the example
of Fig.\ref{overview}. Consider the excitation of the first normal mode for 
the FPU system with $\alpha =1/2$, $\varepsilon =0.01$ and $N=8192$ particles. 
As will be described below, independently of the initial conditions considered, 
we find that, after sufficiently long integration, $J(t)$ always tends to 
stabilize (apart from small fluctuations) to a final value, called hereafter, 
the `equilibrium value' $J_{eq}$. In practice, $J_{eq}$ is numerically 
computed by taking averages of $J(t)$ in moving time windows long enough to 
absorb fluctuations. One notes that eventually the moving average stabilizes 
to a constant value after $t=t_{eq}$. For $t>t_{eq}$, $J(t)$ only exhibits 
rapid and irregular fluctuations around the constant average $J_{eq}$. We 
argue below that this is an indication for a statistical equilibrium state 
reached by the system. The time $t_{eq}$ is hereafter called the 
{\it equilibrium time}.
The {\it normalized} $J$ is defined as: 
\begin{equation}
\tilde{J}(t)=\frac{J(t)-J(0)}{J_{eq}-J(0)}~~.
\end{equation}

In Fig.\ref{overview}  $\tilde{J} $ 
displays a sigmoidal evolution 
from 0 to 1. The detachment of $\tilde{J}$ from zero, which indicates an 
essential departure from quasi-integrable behavior, takes place at a time 
$t_0$. During the time interval $[0,t_0]$ the behavior of $J(t)$ for FPU 
and Toda is nearly indistinguishable. The time $t_0$ is hereafter called 
the {\it time of stability}. At longer times, $\tilde{J}$ tends sigmoidally 
towards the value 1, reached at time $t_{eq}$. 
 
It is crucial to clarify in which sense $t_{eq}$ reflects the time in which
statistical equilibrium has been reached. We give numerical evidence in 
Fig.\ref{eq}, and we theoretically justify below, that the stabilization 
of $J$ to $J_{eq}$ for $t>t_{eq}$ is related to the following: 
(i) the variables $p_n$, $\delta q_n$ decorrelate (Fig.\ref{eq}(b),(f)),  
(ii) the energy spectrum $E_k$ reaches equipartition (Fig.\ref{eq}(d)). 

Regarding (i), Fig.\ref{eq}(b) displays the evolution of the sums 
$\sum p_np_{n+1}$, $\sum \delta q_n\delta q_{n+1}$ for the same orbit as 
in Fig.\ref{overview}. Similar behavior is found for all orbits with initial 
conditions corresponding to localized excitations. For $t<t_0$ both sums 
oscillate, with maxima of the one sum corresponding to minima of the other. 
Between $t_0$ and $t_{eq}$, both sums tend sigmoidally to zero, while, 
beyond $t_{eq}$ they become uncorrelated and both fluctuate irregularly 
around zero. Let us note, in respect, that a correlation sum similar to 
the above was considered by Parisi \cite{parisi}
$$
\Delta (t) = \frac{\overline{\sum_n q_nq_{n+1}} }{ \overline{\sum_n q_n^2}}
$$  
as an accurate and simple observable to be used in specifying the 
equilibrium time. 

We will now show, that property (ii) above is connected with the correlation 
sums $\sum p_np_{n+1}$, $\sum \delta q_n\delta q_{n+1}$ both tending to zero 
for times beyond $t_{eq}$. To this end, for specific energies much smaller 
than unity, $J$ is well approximated by expanding the exponentials in 
(\ref{j1}) in Taylor series in powers of the quantities $p_n$, $\delta q_n$. 
Up to quadratic terms we get (see Appendix 
\ref{analytic}):
\begin{eqnarray} \label{Jrs} 
J \simeq 2  \varepsilon + { 1 \over { 2 N}} \sum_n \Big[ p_n p_{n+1} 
+ \delta q_n \delta q_{n+1}  \Big] ~~.
\end{eqnarray}
Due to Eq.(\ref{Jrs}), 
 the near--constancy of $J$ for $t<t_0$  in Fig.\ref{eq}(a) 
 emerges from the counterbalance of the sums $\sum p_np_{n+1}$ and $\sum \delta q_n\delta q_{n+1}$ in Fig.\ref{eq}(b), 
which oscillate around a non--zero mean with nearly opposite phases. 
For $t>t_{eq}$, however, they both tend to zero.
In fact, as shown in 
Appendix \ref{analytic}, Eq.(\ref{Jrs}) can be recast as a sum over the 
 energy spectrum $E_k$:
\begin{eqnarray} \label{Jnm}
J\simeq 2   \varepsilon  +   { 1 \over {N}} 
\sum_k \cos \big( \frac{\pi k} {N} \big) E_k ~~.
\end{eqnarray}
This indicates that $J$ nearly corresponds to a linear combination of the 
 energies $E_k$, where more weight is given to the energies $E_k$ 
corresponding to low and high frequency modes, and less weight to modes in the 
middle of the spectrum ($k\approx N/2$). Most notably, the sigmoid transition 
of $J$ from the value $J(0)$ to $J_{eq}$ is connected with the evolution of 
the energy spectrum $E_k$ from a localized one (for $t<t_0$) to 
near-equipartition $E_k\simeq\varepsilon$ for times $t>t_{eq}$. Since 
$\sum_{k}\cos(\pi k/N)=0$, one gets $J_{eq} \simeq 2 \varepsilon$, 
implying that $t_{eq}$ can also be interpreted as the `equipartition time'. 
On the other hand, the distribution of the momenta for $t>t_{eq}$ approaches 
a Gaussian with dispersion $\sigma_p^2=2\varepsilon$ (cf. Fig.\ref{eq}(e) 
and (f)). This is not far from the Gibbs measure for an equilibrium state 
with Hamiltonian (\ref{fpuham}). Thus, while energy equipartition alone does 
not necessarily imply an equilibrium state (see also numerical simulations 
below), in the above example energy equipartition is linked to and appears 
at the same timescale as the decorrelation between the phase space variables. 

\begin{figure}
\centering
\includegraphics[scale=0.45 ]{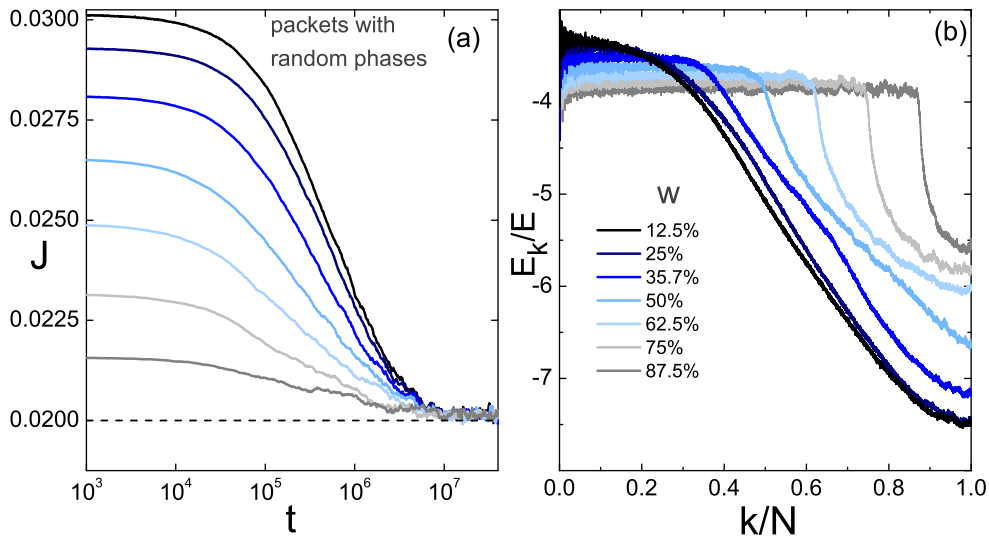} 
\caption{Initial excitation of the modes $0\leq k/N \leq w$ with random initial phases 
(as indicated in the figure) for $N=8192$, 
$\alpha =1/2$ and $ \varepsilon =0.01$: (a) evolution of $J$, 
(b) Time--averaged energy spectra during the metastable state (for $t<t_0$).}  
\label{rp}
\end{figure}

It is noteworthy that approximating formulas analogous to Eq.(\ref{Jnm}) 
can be derived relating {\it all} the Toda integrals with linear combinations 
of the spectral energies $E_k$. Thus, a behavior analogous to the one 
found above for $J$ is expected to hold for all Toda integrals. This subject
is currently under investigation.

Finally, we remark that the evolution of $\tilde{J}$ in Fig.\ref{overview}
is nearly indistinguishable from the evolution of the {\it tail energy}
\footnote{
A  idea similar to the tail energy for generic packet excitations 
can be traced back to reference \cite{PP05}.}
\begin{equation}
\eta =2\sum_{k\geq N/2} E_k/E
\end{equation}
which has been proposed as an efficient observable for distinguishing 
stages of dynamics in the FPU problem (see \cite{dresden,benettin1}). An 
additional method based on the times at which the FPU trajectories 
intersect an `equilibrium manifold' \cite{flachlast} yields results 
similar to those of \cite{dresden} found using the tail energy 
evolution. We find that the normalized Toda integral $\tilde{J}$ 
and the tail energy practically coincide for trajectories with 
initial low-mode excitations. However, as discussed below, $\tilde{J}$ 
is of use also in more generic initial conditions, in which the 
tail energy cannot be used.

\section{Numerical Experiments}
\label{le}
In this section we numerically investigate the stability, diffusion, and 
approach to equilibrium for FPU trajectories with various types of initial 
conditions, using as a probe the evolution of $J$.  

\begin{figure}
\centering
\includegraphics[width=0.6\textwidth]{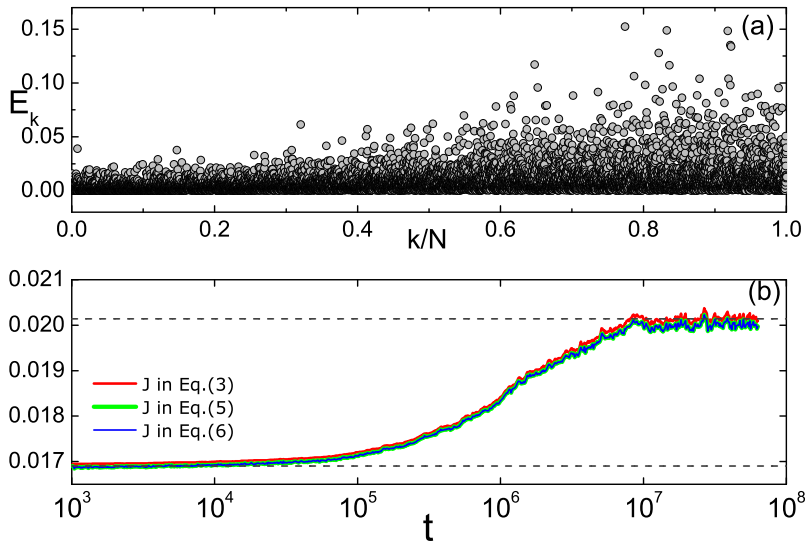}
\caption{ FPU with $N=8192$, $\alpha =1/2$ and $\varepsilon =0.01$. 
(a) Example of the harmonic energy spectrum $E_k$ for an initial condition 
corresponding to random (non-Gaussian) positions and momenta. (b) Evolution 
of $J$ (red) and its two quadratic approximations according to Eqs.(\ref{Jrs}) 
(green) and (\ref{Jnm}) (blue).} 
\label{deloc}
\end{figure}
\begin{figure}
\centering
\includegraphics[width=0.6\textwidth]{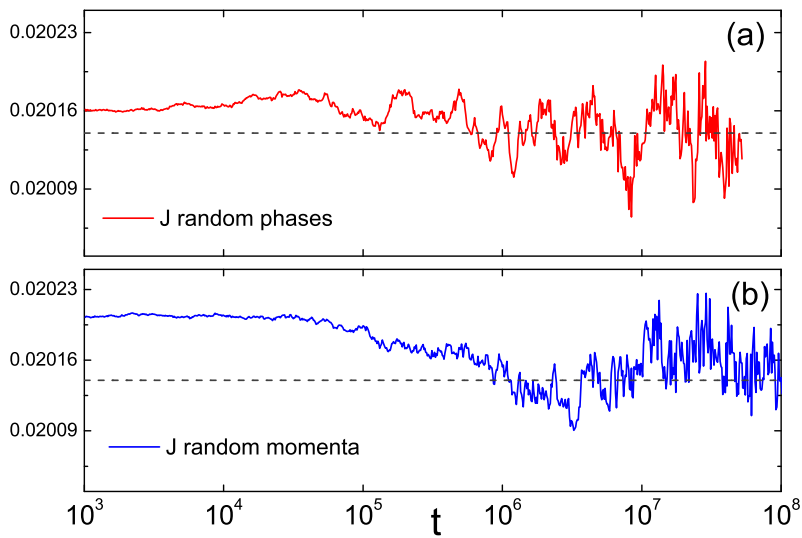}\\
\caption{ Evolution of $J$ for two types of initial conditions in the FPU 
system with $N=8192$, $\alpha =1/2$ and $\varepsilon =0.01$. (a) Energies 
starting at equipartition $E_k(0)=\varepsilon$ with randomly chosen 
initial phases for the normal mode variables, and (b) random positions with 
uniform distribution (see text). Both panels show the evolution of the 
mean $J$ from 10 realizations per class of initial conditions. According to 
the formulas (\ref{Jrs}) and (\ref{Jnm}) in both cases we have $J(0)\simeq 
J_{eq}$ for each individual trajectory.}
\label{ap1}
\end{figure}

\subsection{Classes of initial conditions and the evolution of J}
\label{inicon}
We cluster different types of initial conditions into three main categories 
referred to below as: (i) with `localized energy spectrum', (ii) with 
`delocalized energy spectrum', and (iii) `close to equipartition'.\\
\\
{\it (i) Localized energy spectrum:} Case (i) contains FPU trajectories 
with initial conditions corresponding to single--site excitations, and 
packets of modes with coherent or random phases. The initial conditions 
are controlled through the initial values assigned to the normal mode 
variables $(Q_k,P_k)$, $k=1,\ldots, N-1$. To an initial energy $E_k(0)$ 
assigned to mode $k$ there corresponds a family of possible initial 
conditions with $Q_k(0)=A_k\sin(\phi_k)$, $P_k(0)=\omega_kA_k\cos(\phi_k)$ 
where $A_k=(2E_k(0)/\omega_k)^{1/2}$ and $0\leq\phi_k<2\pi$. A single--site 
excitation is obtained by setting $A_k=0$ for all $k$ except one $k=k_0$. 
Here we consider the case $k_0=1$. A `packet of modes' excitation of 
percentage $w$ corresponds to setting $E_k(0)=N\varepsilon/w$ for all $k$ 
in $1\leq k\leq [wN]$ with $0<w\leq N$ and $E_k(0)=0$ otherwise. In this 
case, we can have initial phases `coherent' ($\phi_k=const$, here 
$\phi_k=0$), or `random' (here $\phi_k$ is chosen from a uniform 
distribution in $[0,2\pi)$). 

Figure \ref{rp}(a) shows the evolution of $J$ along FPU trajectories 
with initial conditions corresponding to packets of various sizes (values 
of $w$ as denoted in the figure) and random phases, while the specific 
energy is kept fixed at $\varepsilon=0.01$. The $12.5 \%$ packet (black) 
gives rise initially to a strongly localized energy spectrum which is nearly 
flat among the packet modes and develops an exponential tail for the rest of 
the modes, as shown in Fig.\ref{rp}(b). By progressively increasing the width 
of the packet (from $12.5 \%$ to $87.5 \%$), different localization patterns 
emerge in Fig.\ref{rp}(b). However, the evolution of $J$ in Fig.\ref{rp}(a) 
shows that equilibrium is reached at nearly equal times $t_{eq}$ for all 
these trajectories. The dependence of both the `stability time' $t_0$ and 
the equilibrium time $t_{eq}$ on $N$ and $\varepsilon$ is discussed in 
subsection \ref{eqtimes}.

{\it (ii) Delocalized energy spectrum:} case (ii) refers to 
non--equilibrium initial conditions with substantial power in the whole 
normal mode energy spectrum $E_k$. As a basic example, we consider random 
initial positions and momenta $(q_n,p_n)$ extracted from a uniform 
distribution in the intervals 
$-p_\varepsilon\leq p_n\leq p_\varepsilon$, 
$-q_\varepsilon\leq q_n\leq q_\varepsilon$, 
with $p_\varepsilon=q_\varepsilon$ tuned numerically so that the total 
energy takes a selected value $E=N\varepsilon$. Fig.\ref{deloc}(a) shows 
the initial energy spectrum $E_k$ for such a choice of initial conditions, 
leading to specific energy $\varepsilon=0.02$. At $t=0$ the system 
appears as not being too far from energy equipartition, 
however, such initial conditions are not only far from the Gibbs measure
but need longer equilibrium times than low--mode excitations. As shown 
in Fig.\ref{deloc}(b) for a single realization, 
$J(t)$ exhibits a similar sigmoidal evolution to the case (i). The approximations (\ref{Jrs}) and (\ref{Jnm}) 
yield curves nearly indistinguishable from the exact curve $J(t)$. Thus, 
the sigmoidal evolution is related to the slight redistribution of energies 
taking place as the system evolves to equilibrium. These are hard to 
distinguish using measures based directly on the spectrum $E_k$, as for 
example, the spectral entropy $S$; see \cite{Livi}). However $J(t)$ measures 
with accuracy the time to equilibrium $t_{eq}\approx 10^7$. Remarkably, 
$t_{eq}$ for random initial positions and momenta turns to be of the same 
order as for the packet initial conditions, and actually somewhat {\it larger} 
than the times $t_{eq}$ found for any `{\` a} la Fermi' type of initial 
condition. This implies that stickiness to the Toda dynamics is a generic 
property {\it not} restricted to trajectories with localized initial 
energy spectrum.

{\it (iii) Initial conditions close to equipartition:} here we consider two 
subcases of initial conditions, namely far from or close to the Gibbs 
distribution $f(\mathbf{q},\mathbf{p})\propto e^{-\beta H_{FPU}(\mathbf{q},
\mathbf{p})}$. In the first subcase, we consider initial conditions called 
hereafter `random momenta', in which we simply set $q_n=0$, and $p_n$ randomly 
chosen from a uniform distribution in the interval $-p_\varepsilon\leq p_n\leq 
p_\varepsilon$, with $p_\varepsilon=\sqrt{6\varepsilon}$. One obtains 
$<P_k^2>=<p_n^2>=2\varepsilon$, and hence a spectrum $E_k$ with 
$<E_k>=\varepsilon$. As an alternative, we fix the normal mode variables 
$P_k=\omega_kA_k\cos\phi_k$, $Q_k=A_k\sin\phi_k$, with $A_k=
(2\varepsilon/\omega_k^2)^{1/2}$ and $\phi_k$ randomly chosen with uniform 
distribution in the interval $[0,2\pi)$. Both these types of initial 
condition lead to spectra close to equipartition, but far from the 
distribution function associated with statistical equilibrium. Instead, 
in the second subcase, hereafter called `close to equilibrium', as in the 
works \cite{benettin2,benettin3,carati,carati2} we approximate a Gibbs distribution function $f$ for the normal mode 
variables $(Q_k,P_k)$ by considering only the quadratic part of the Hamiltonian 
$H_{FPU}$. The function $f$ becomes a Gaussian
\begin{equation}\label{dfgauss}
f\propto\exp\left(-{\beta\over 2}\sum_{k=1}^N(P_k^2 +\omega_k^2Q_k^2)\right)~.
\end{equation}
Setting $\beta = \varepsilon^{-1}$ ensures the specific energy is equal to 
$\varepsilon$ for any random realization in the above distribution.  

\begin{figure}
\centering
\includegraphics[scale=0.3 ]{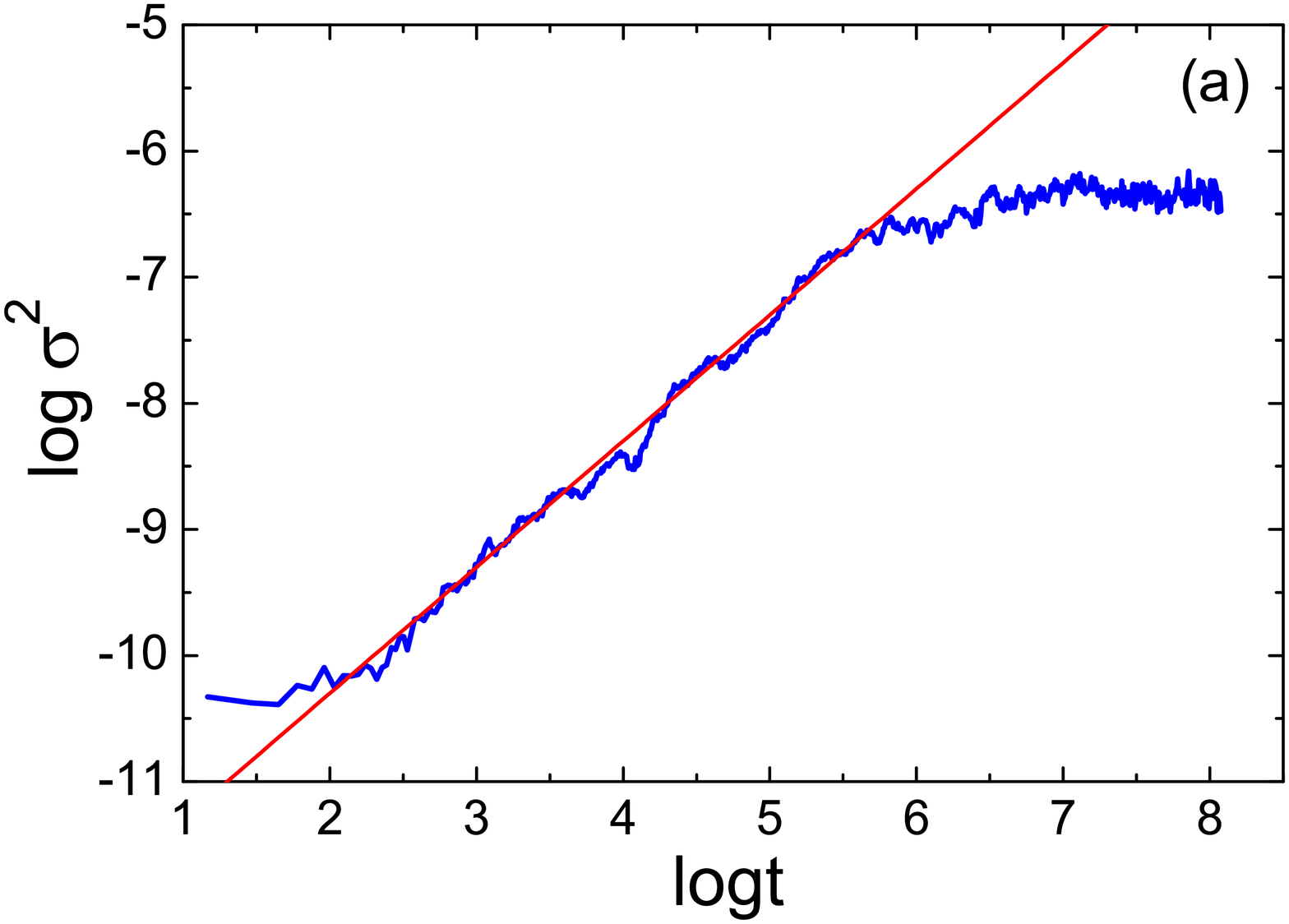} 
\includegraphics[scale=0.3 ]{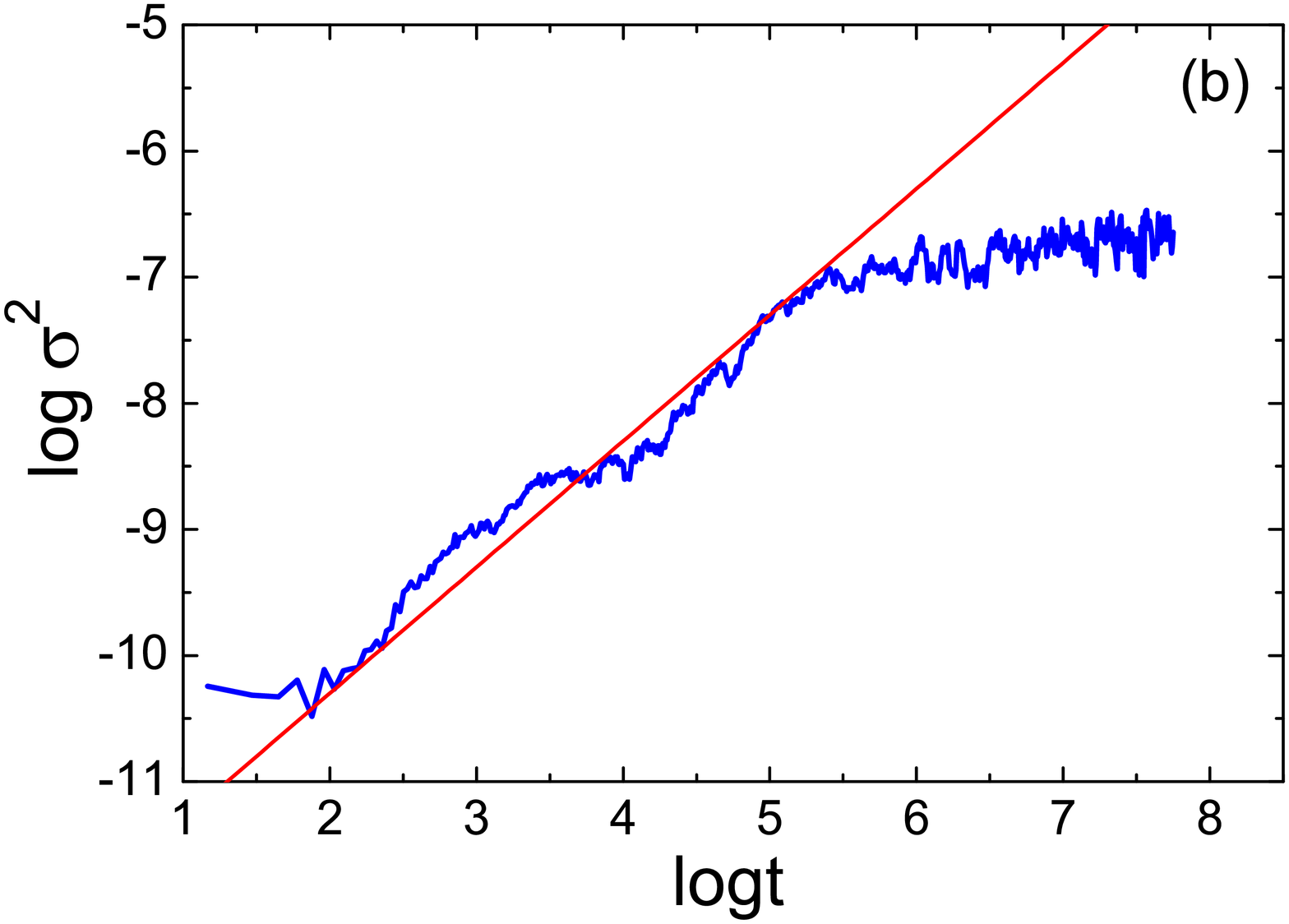}
\caption{The dispersion $\sigma^2_J(t)$ vs. time in logarithmic scale, 
measuring the diffusion spread transversally to the integral surface defined by 
the Toda integral $J$. (a) Results from 30 trajectories in the `random 
initial momenta' case for $\varepsilon=0.01$. (b) Same but for 
the `close to equilibrium' case.}
\label{difsig}
\end{figure}

One key feature of the numerical experiments with all the above initial 
conditions is that $J(t)$ (for single trajectories, or averaged over 
trajectories with different realizations of the initial conditions) 
exhibits no systematic change of its value (e.g. of sigmoid form) allowing 
to define a timescale to equilibrium as in the case of the experiments 
in class (i) and (ii). Fig.\ref{ap1}(a) and (b) shows two examples of the 
evolution of the average $J(t)$ for ten realizations of trajectories with 
initial conditions belonging to the `random phase' and `random momenta' 
subclass. Despite the absence of sigmoid evolution, one may still argue 
that for initial conditions far from equilibrium, a certain time is required 
for these trajectories to drift substantially in phase space so as to 
approach a state of equilibrium. Such an approach cannot be detected by 
the evolution of the energy spectra $E_k$, since the latter are set from 
the start close to energy equipartition. Nevertheless, the form of the 
curves in Fig.\ref{ap1} suggests the trajectories undergo diffusion 
in the direction normal to the integral surfaces defined by the Toda 
integrals, or at least the first one. Such diffusion can be detected 
by measuring the evolution of the dispersion in the values of $J(t)$ 
over $M$ trajectories
\begin{equation}\label{dispj}
\sigma_J^2(t) = {1\over M}\sum_{l=1}^M \left(J_l(t)-J_l(0)-\mu_J(t)\right)^2 
\end{equation}
where $\mu_J(t) = {1\over M}\sum_{l=1}^M \left(J_l(t)-J_l(0)\right)$. 
Subtracting the initial value $J_l(0)$ is necessary to absorb the 
spreading in the initial values of $J_l$ due to the random generator of 
initial conditions. Fig.\ref{difsig}a Shows $\sigma^2(t)$ as a function 
of $t$ for a set of 30 trajectories with `random initial momenta'. 
The solid line has logarithmic slope equal to 1, indicating a normal 
diffusion law $\sigma^2\propto t$. Notice that the diffusion stops 
after a time $t\sim 10^6$. The spreading in the values of $J$ after 
this time has measure $O(\varepsilon^2)$, indicating that the trajectories 
have spread over parts of the phase space covering the whole possible 
range of values of $J$ consistent with a fixed specific energy equal to 
$\varepsilon=0.01$. 

On the other hand, it is really remarkable that a similar spreading is 
found for initial conditions very close to the Gibbs measure, as exemplified 
in Fig.\ref{difsig}(b). Now, the initial conditions are chosen by the 
distribution function of Eq.(\ref{dfgauss}), thus they represent a state 
as close as possible to statistical equilibrium already at the starting 
point of the simulation. Yet, we observe that the underlying integrable 
dynamics of the associated Toda model leaves its traces in this case too, 
since the trajectories diffuse transversally to the integral surface of the 
integral $J$ with a very slow speed, comparable to the one in initial 
conditions far from equilibrium. Measuring this speed at various energy 
levels, as well as for different Toda integrals represents a challenging 
numerical task, since it requires  many 
trajectory realizations per parameter set considered in order to ensure 
a good statistics (setting $M=30$ in the above experiments is marginal 
in this respect). Instead, the speed of approach to equilibrium is measured 
more easily via the sigmoid curves in experiments of classes (i) and (ii), 
a computation to which we now turn our attention.

\subsection{Intensive quantities and a single timescale}
\label{eqtimes}

\begin{figure}
\centering
\includegraphics[scale=0.4]{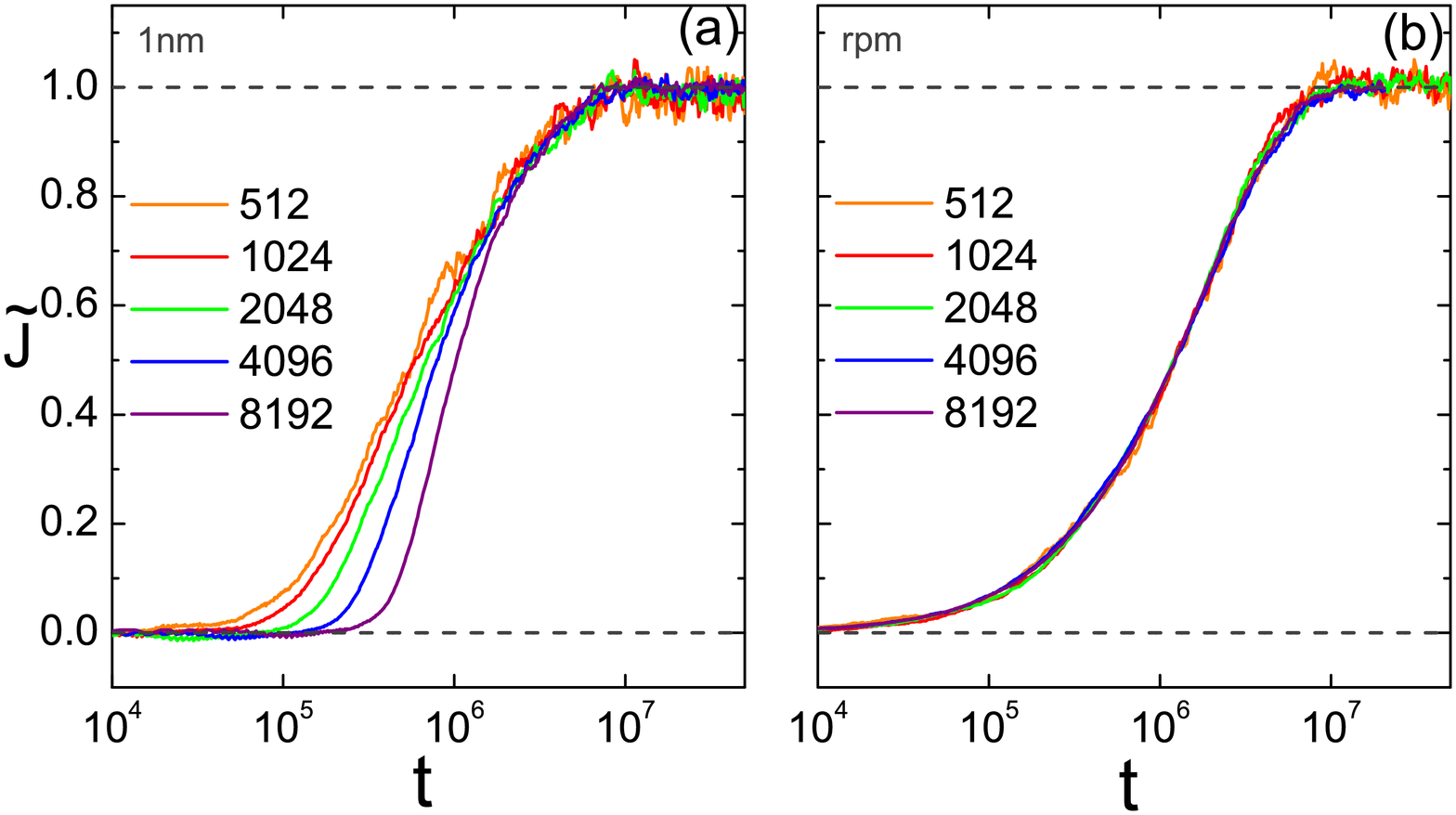}
\includegraphics[scale=0.4 ]{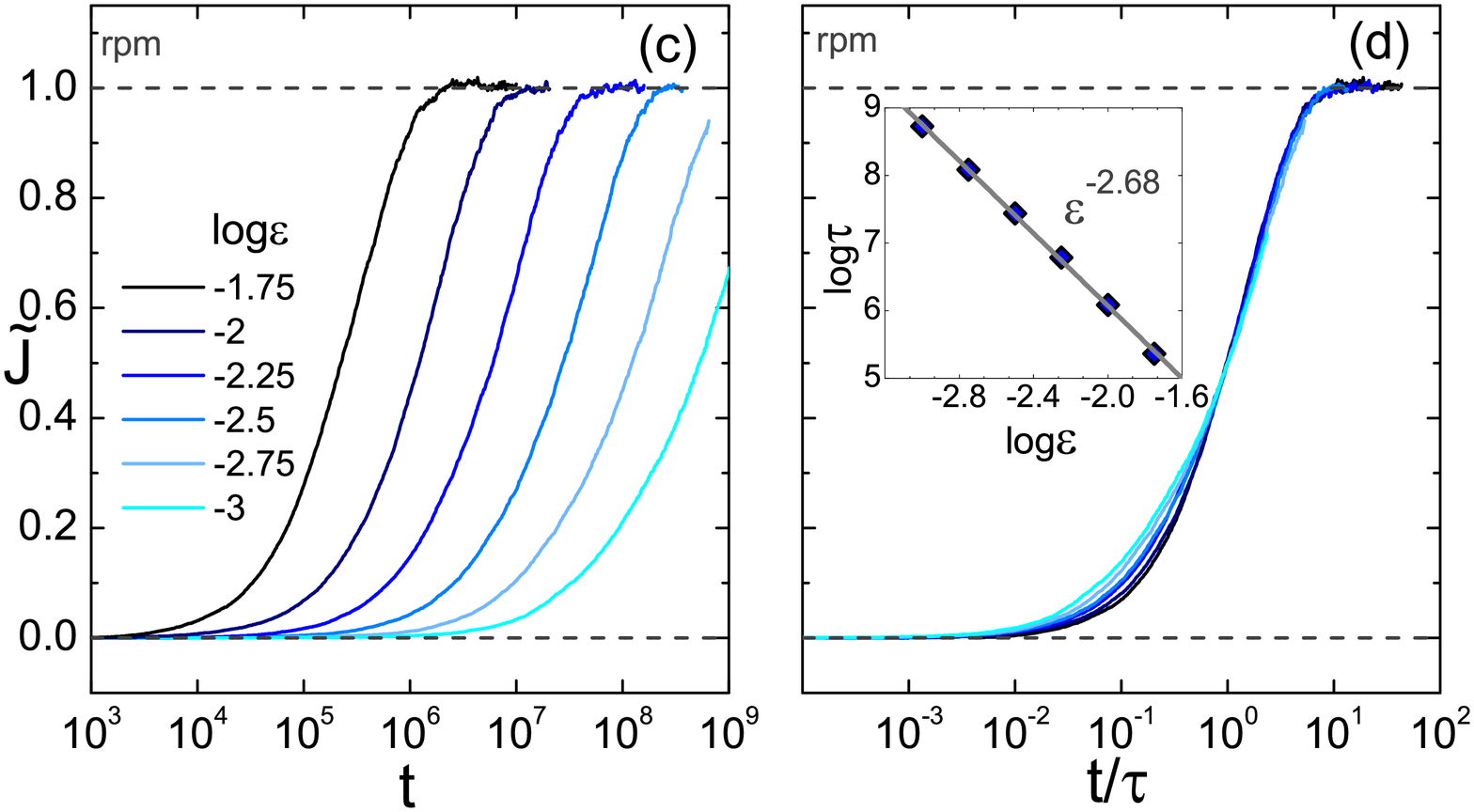} 
\caption{ Upper panels: The normalized Toda integral $\tilde{J} $ along the 
FPU dynamics for $\alpha =1/2$ and $\varepsilon =0.01$ and various $N$ values 
for: (a) the excitation of the first normal mode, and (b) random initial positions and momenta. 
Lower panels:(c) The sigmoid evolution of $J$ for random positions and momenta at various energies. (d) By rescaling 
the time as $\tau \sim \varepsilon ^{-2.7}$, all sigmoid curves $J$ cluster in one: 
$J(t/\tau )$.
\label{ext}  }
\end{figure}

We first investigate for which types of initial conditions the behavior of 
$J(t)$ and associated timescales (of `stability' or 'approach to equilibrium', 
see section 2) exhibit a dependence on $N$. It turns out that only 
{\it random initial data lead to $N$--independent results}. For example, 
Fig.\ref{ext}(a) shows the sigmoid curves when considering the excitation 
of the first normal mode (case (i)), with $N$ ranging from 512 to 8192. 
The curves are distinguished, in particular as regards the `stability' 
times characterized by the onset of the rising part of the sigmoid 
evolution, even if the equilibrium times are similar for all $N$. 
In comparison, Fig.\ref{ext}(b) shows the same computation but for initial 
conditions extracted by random positions and momenta (case (ii)). Superposing 
the sigmoid curves for various $N$ shows a near-coincidence at all times.

We note that a similar behavior to Fig.\ref{ext}(b) holds for initial data as those of 
Fig.\ref{rp}.
Thus, packet excitations with random phases exhibit $N$--independence in 
the timescales associated with the evolution of $\tilde{J}(t)$. This 
is in agreement with the results first reported in \cite{Livi2}, where 
the authors aim to unify `incompatible' earlier findings in the FPU 
literature (see \cite{Livi2} and references therein) on the dependence of 
various scaling laws on the energy $E$, when exciting packets of modes with 
coherent phases, or the specific energy $\varepsilon$, when the initial phases 
are random. It is also in agreement with the work \cite{tori3}, 
where it is found that extensive packets with coherent phases give rise 
to exponentially localized energy spectra of the form $E_k \sim  
e^{-\sigma k/N}$, with $\sigma \sim (\alpha ^2 E)^{-d}$, $d>0$, i.e., 
depending on $E$ rather than $\varepsilon$. 

Exploiting the property of $N$--independence, we can characterize the 
timescales involved in the evolution of $J(t)$ for classes of trajectories 
based on some form of random initial data. Computing $J(t)$ allows to obtain 
a good estimate of the `stability time', from the size of the initial plateau 
of the corresponding sigmoid curve, and the time of approach to equilibrium, 
when the second plateau is reached. Due to the numerical indications for 
$N$--independence one may reasonably argue that the results hold in the 
thermodynamic limit ($N\rightarrow\infty$ keeping $\varepsilon$ constant). 
Focusing, as an example, on trajectories with random initial positions and 
momenta, the sigmoid curves for the normalized Toda integral $\tilde{J}(t)$ 
are reproduced in Fig.\ref{ext}(c). With an appropriate time--shift 
$\tau \sim \varepsilon ^{-2.68}$ (inset), those curves fall one onto 
the other as shown in Fig.\ref{ext}(d). We typically 
find $\tau \sim \varepsilon ^{-a}$ beyond a crossover specific energy 
(see below) for some $a>0$ depending  on the type of initial 
conditions. Since the shift in the sigmoid curves takes place in a 
logarithmic time axis, the above facts imply that $\tau$ allows to 
characterize the scalings with $\varepsilon$ of {\it both} the initial 
`stability time' and the `time to equilibrium'. One has 
\begin{eqnarray} \label{ttt} 
t_0(N,\varepsilon ) = t_0(\varepsilon ) \simeq c_1 \tau \nonumber\\
t_{eq}(N,\varepsilon ) = t_{eq}(\varepsilon ) \simeq c_2 \tau \nonumber
\end{eqnarray} 
with $c_1<1<c_2$. 
 At practical level, the possibility to 
extrapolate scaling laws from the `stability' to the `approach to equilibrium' 
phase allows to translate results found for $t_0$ to analogous results for 
$t_{eq}$, even when $t_{eq}$ is much larger than $t_0$ and hence hard to 
compute by direct numerical experiments. One has to compute in respect  
the initial part of the curve $J(t)$, up to the detachment from the first 
plateau. We also note that the numerical complexity for computing $J$ is 
$O(N)$, which is much better than the $O(N^2)$ complexity of any computation 
requiring knowledge of the evolution of the energy spectra $E_k$. 

\begin{figure}
\centering
\includegraphics[scale=0.4 ]{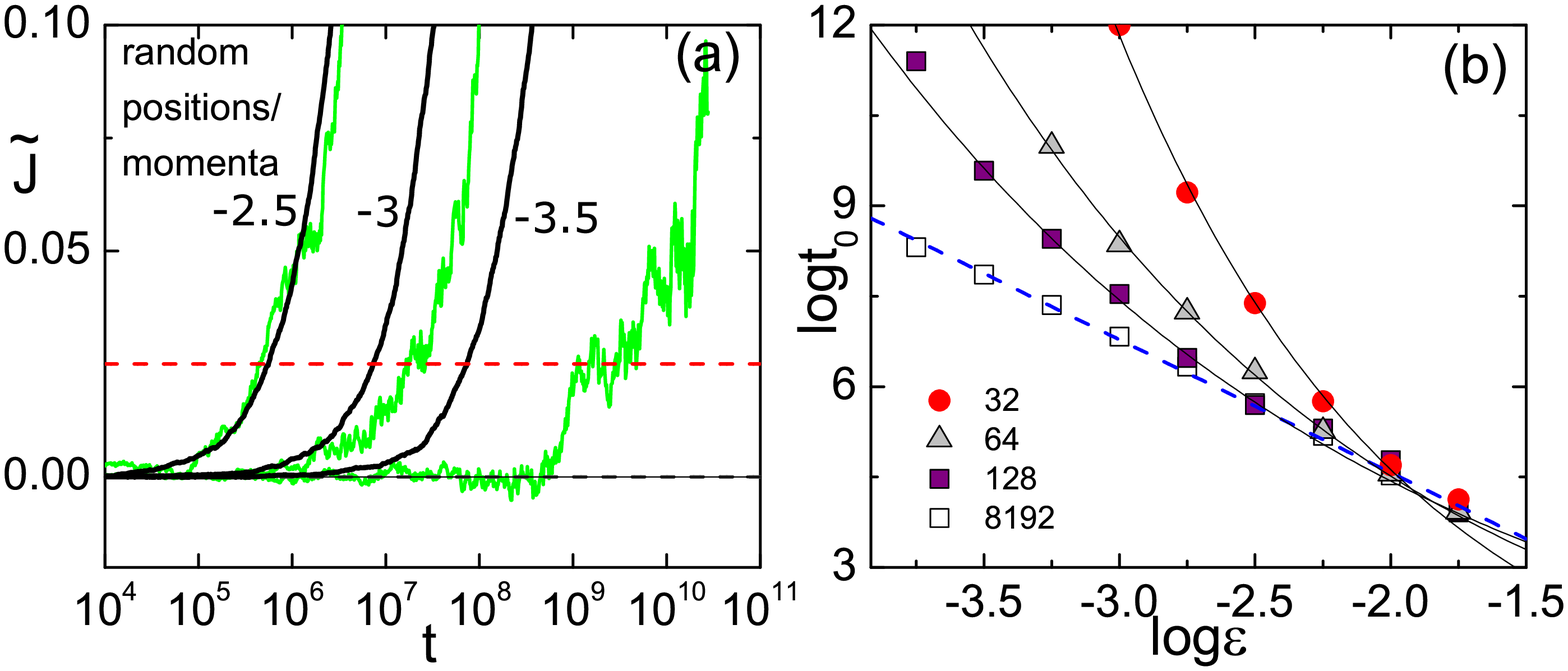}
\includegraphics[scale=0.4 ]{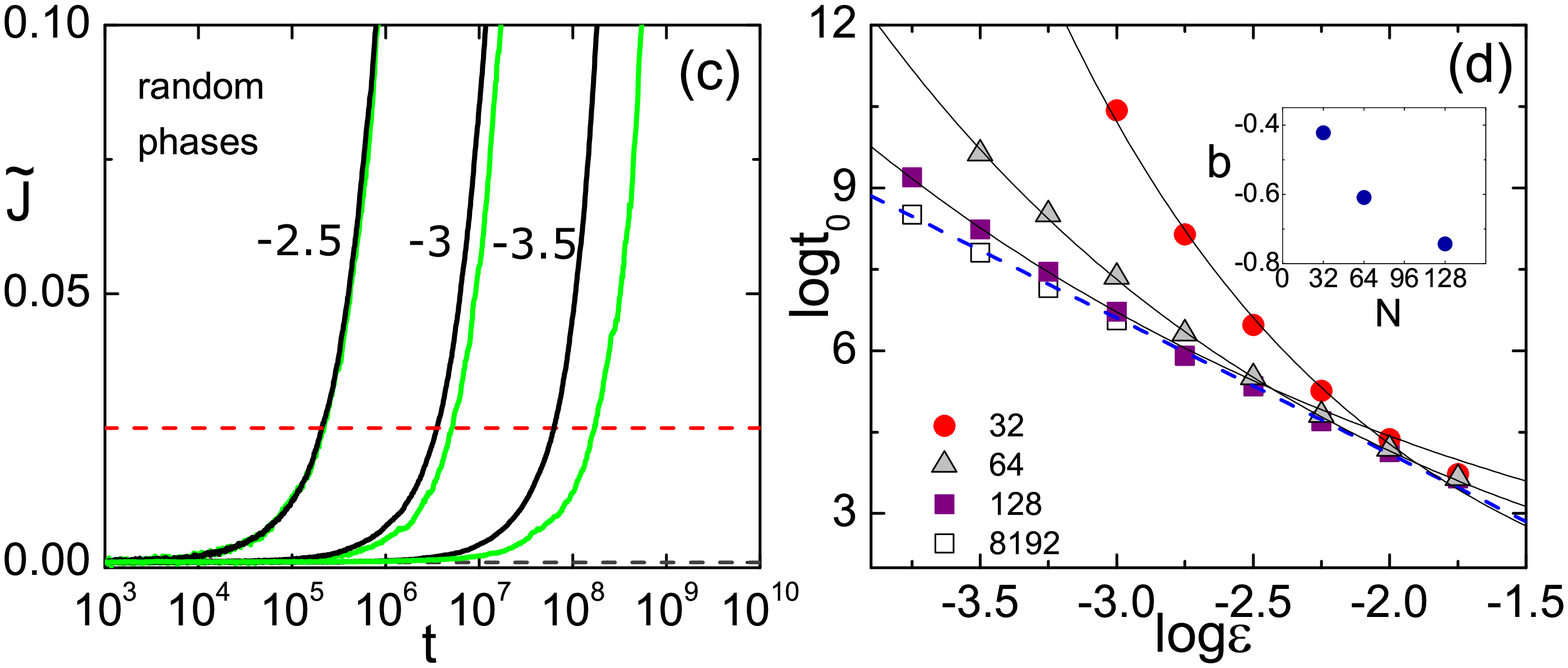} \\
\includegraphics[scale=0.4 ]{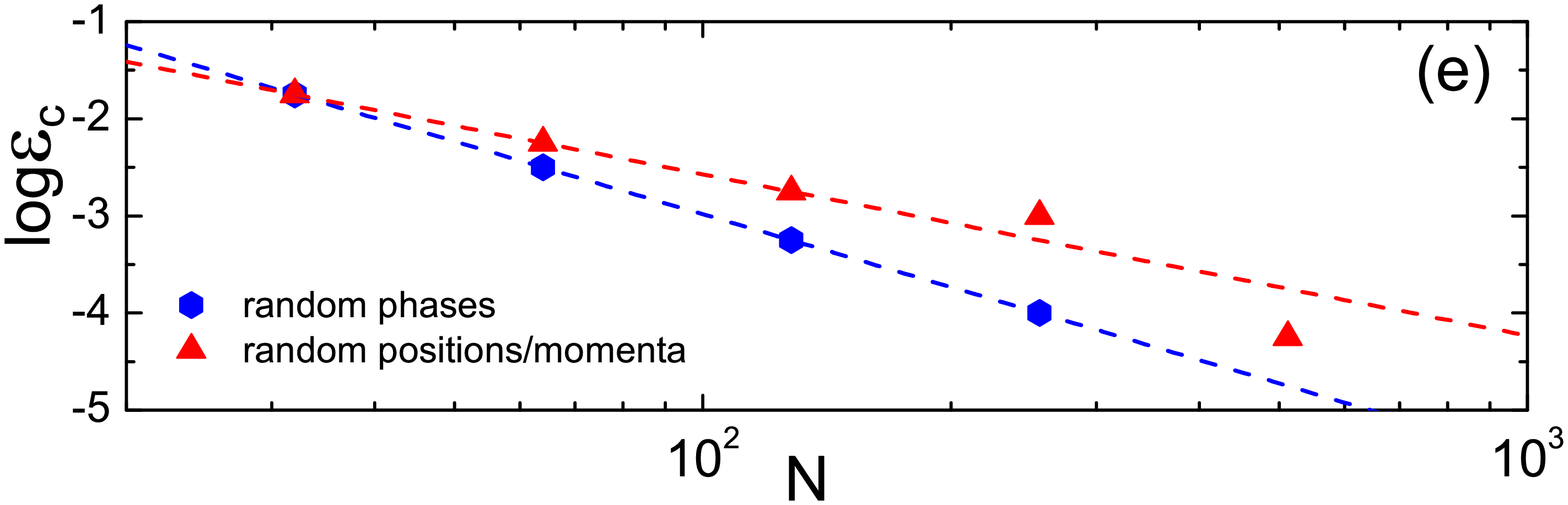} 
\caption{  (a) Random initial positions and momenta: 
$\tilde{J}$ for the exemplary $N=8192$ curve (black) 
and its disagreement with $N=128$ (green curves), which is evident below 
$\varepsilon_c =10^{-2.75}$.  (b) Stability times for the case (a): $N=8192$ (power-law) 
and the gradual exponential divergence for $N=128$, $64$ and $32$ particles. 
(c) and (d): same as (a) and (b) respectively, when the first $12.5 \%$ of 
modes are excited with random phases. (e) The specific energy crossover of 
the two cases versus $N$ decays to zero as $N^{-1.5}$ ((a),(b) case with red triangles) 
 and  as $N^{-2.5}$ ((c),(d) case with blue polygons).}
\label{128}
\end{figure}

All the above results hold asymptotically for $N$ sufficiently large. 
Instead, for $N$ small we may find deviations from the law  $\tau \sim \varepsilon ^{-a}$
which are $N$--dependent. In particular, we find transient exponential laws 
$\tau \sim  \exp \varepsilon ^{-b(N)}$, $b(N)>0$ up to a crossover specific energy $\varepsilon _c(N)$. 
However, we can make use of $J$'s asymptotic independence in order to locate the 
specific energy crossover $\varepsilon _c(N)$. 
The numerical procedure is the following: 
the sigmoid curves $\tilde{J}(t)$ or $\tilde{J}(t/\tau )$ (as in 
Fig.\ref{ext}(d)) serve as exemplary curves which determine the power-law dependence 
on $\varepsilon $ of the timescales $t_0$ and $t_{eq}$. These curves are $N$--independent for 
random--based initial conditions. Starting with small system sizes $N$ and by 
lowering the energy, one eventually encounters the crossover: {the law 
describing these timescales changes from power to exponential} and a 
dependence on $N$ emerges. Practically, this means that for $\varepsilon
<\varepsilon_c$ the $\tilde{J}$--curves  do not match with their exemplary 
counterparts, as in the example of Fig.\ref{128}.

Two sets of three sigmoid 
curves are shown in Fig.\ref{128}(a) (experiments with random initial positions 
and momenta) for decreasing energies with constant logarithmic step, namely 
$\log \varepsilon =-2.5$, $-3$, $-3.5$ and $N=8192$ (black set) or $N=128$ 
(green set). Only the initial parts of the $\tilde{J}(t)$--curves are shown, up to 
 the value $0.1$. At $\log \varepsilon =-2.5$ the 
curves for the two values of $N$ nearly coincide, while a mismatch appears  
at $\log\varepsilon_c =-2.75$, clearly becoming larger by further lowering 
the energy. Fig.\ref{128}(b) explains this mismatching in terms of
the stability time $t_0$, defined as the times when the curve $\tilde{J}(t)$ 
first crosses the value $0.025$. Inspecting Fig.\ref{128}(b), the delineation 
of $t_0$ to the $N$--independent final power-law (dashed line, obtained by 
fitting the data for $N=8192$) occurs at {\it smaller} crossover specific 
energies $\varepsilon_c$ as $N$ increases. Repeating the study for the 
initial excitation of a $12.5 \%$ packet of modes with random phases
(Fig.\ref{128}(c) and (d)) find the same trend of $\varepsilon_c$ with 
$N$, with slightly different exponents. Notice that, for energies smaller 
than the crossover, one cannot always safely distinguish between an 
exponential law or a power law with steeper exponent than the exponent 
of the $N$--independent profile. For example, in Fig.\ref{128}(d) the $N=128$ 
case could be fairly well fitted by a steeper power-law, in accordance 
with the interpretation via the `six-wave resonant interactions'  in \cite{onorato}. 
Leaving open such questions, we here emphasize the utility of observing 
$\tilde{J}(t)$ for answering them, as well as for characterizing the 
asymptotic regime, which indicates that $\varepsilon _c\rightarrow 0$ 
as $N\rightarrow \infty$: the latter fact suffices to exclude 
exponentially-long lasting deviations from the equilibrium state at the 
thermodynamic limit ( Fig.\ref{128}(e)). 

\begin{figure}
\centering
\includegraphics[scale=0.3 ]{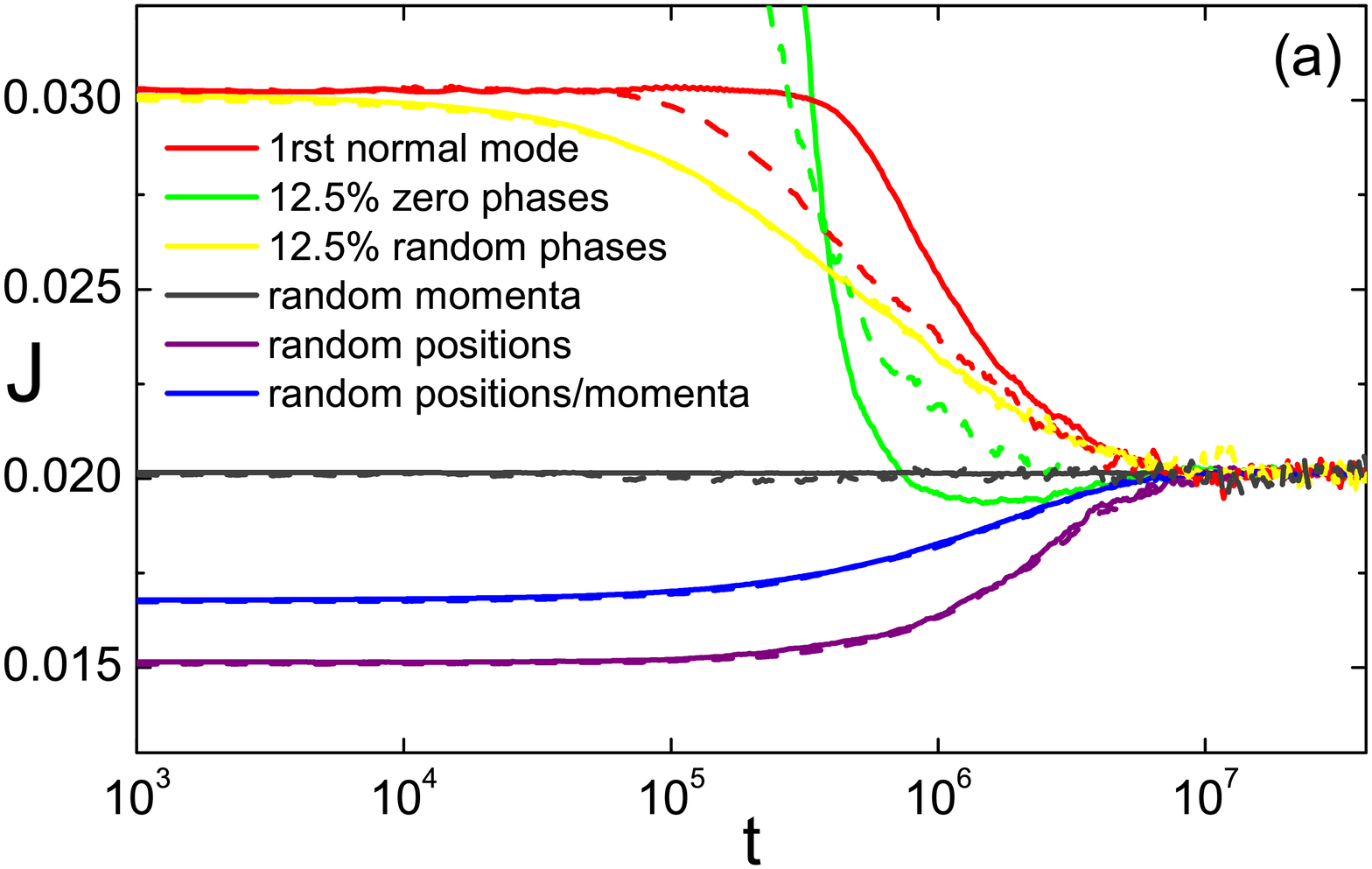} 
\includegraphics[scale=0.3 ]{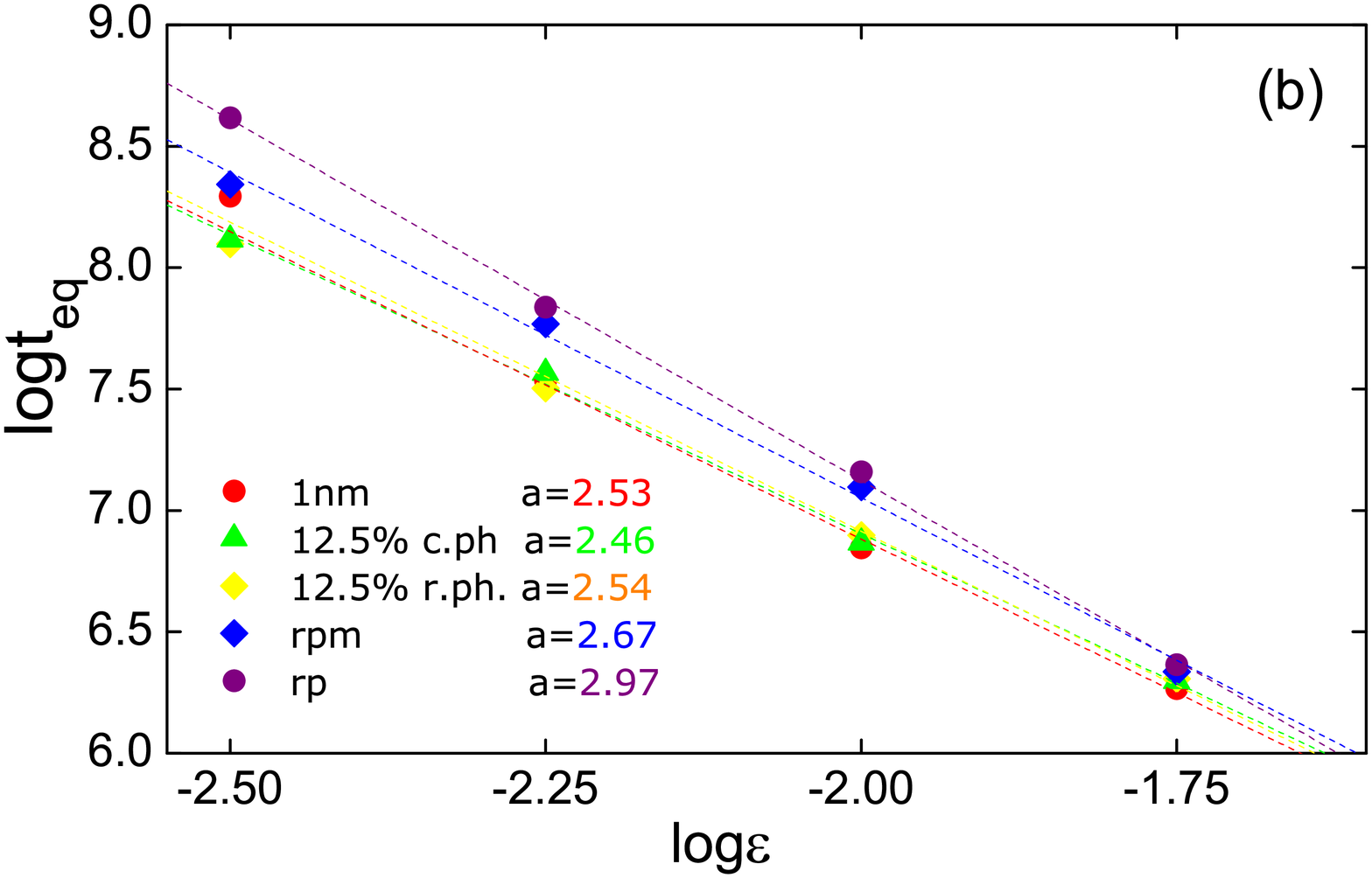}
\caption{(a) Six types of initial conditions, lead to different sigmoid 
curves for $J(t)$. Continuous lines: $N=8192$, dashed line: $N=1024$.
It turns out that the first normal mode and the packets with coherent phases
are non-extensive.  (b) Equilibrium times versus the specific energy, measured 
for the FPU system with $N=8192$, $\alpha =1/2$ and for 5 classes of initial 
conditions.
\label{spectra}  }
\end{figure}
Finally, we examine how the scaling laws on equilibrium times are affected 
by particular choices of initial conditions. To this end we consider six 
different types of initial conditions for the FPU system with $\alpha =1/2$ 
and $ \varepsilon =0.01$, exciting: (1) the first normal mode, (2) the $12.5\%$ 
lowermost frequency packet of modes with zero initial phases (as in 
\cite{tori1,tori2,tori3}), (3) the $12.5\%$ lowermost--frequency packet of 
modes with random initial phases (as in \cite{Livi2}), (4) random initial 
momenta, (5) random initial positions and finally, (6) random initial 
positions/momenta. The corresponding sigmoid curves are displayed in 
Fig.\ref{spectra}(a), with solid lines for $N=8192$ and dashed lines for 
$N=1024$ ($\varepsilon =0.01$ in all cases). We immediately note the 
$N$--dependence of the behavior of the sigmoid curves for (1) and (2).  
Also, in (4) the spectrum $E_k$ exhibits equipartition already at $t=0$, 
hence $J(0)=J_{eq}$. On the other hand, in (5) and (6) $J$ starts below 
$J_{eq}$, since the energy spectrum somewhat favors high modes (see 
Eq.(\ref{Jnm})). Finally, all curves equilibrate at $J_{eq} \approx 0.02015$, 
i.e. very close to $0.02$, as predicted by Eq.(\ref{Jnm}) for all $E_k$ 
equal.

Using now the numerical procedure described above, we calculate the 
equilibrium times $t_{eq}$ for five out of the six types of initial conditions 
in Fig.\ref{spectra}(a) and for $N=8192$ (we exclude (4) in which 
$J(0)=J_{eq}$). As Fig.\ref{spectra}(b) shows, in all cases we find 
$t_{eq}\sim\varepsilon^{-a}$, with the exponent $a$ varying between 2.5 
(for the low--frequency mode excitations), to 3 (random positions and/or 
momenta). To within numerical uncertainties, one is tempted to conclude 
that yielding, in the initial data, more power to the {\it high-frequency} 
part of the mode spectrum results in steeper laws $t_{eq}\propto
\varepsilon^{-a}$, i.e., trajectories more stable against the approach 
to equilibrium. We leave open the question of the associated scalings 
when only high modes are excited, a case not largely considered so far 
in literature.

\section{Conclusion}
We examined the role of the first Toda integral $J(q,p)$ as an indicator
of the evolution, and crossing of various `stages of dynamics' for FPU
trajectories. This is accomplished by computing the time variations of the
`normalized' Toda integral $\tilde{J}$ (section 2) when an FPU trajectory
$(q(t),p(t))$ is substituted within $J(q,p)$. Our main conclusions are
summarized below.


Despite its apparent simplicity, $J(t)$ proves to be a very efficient
indicator allowing to clearly distinguish stages of FPU dynamics for a
wide class of initial conditions. In particular, it allows to clearly
define two times, called the `time of stability' $t_0$, and the `time
to equilibrium' $t_{eq}$. For times $t<t_0$ the FPU trajectories remain
extremely close to the original integral surface $J(q_0,p_0)=const$.
Thus, $t_0$ can be interpreted as a time of stickiness to the original
Toda integral surface. On the other hand, $t_{eq}$ marks the approach
of the system to equilibrium.

For classes of initial conditions based on random initial data (in
position or momenta, or the phases for normal mode variables),
as long as the initial energy spectrum $E_k$ is not in equipartition,
the times $t_0$ and $t_{eq}$ are connected via a `sigmoid' evolution
of the curves $\tilde{J}(t)$. For large $N$, the graph of $\tilde{J}(t)$
becomes $N$--independent, so that we obtain one representative 
sigmoid curve $J(t/ \tau )$ described by one time-scale 
$\tau \sim  \varepsilon ^{-a}$ with exponents $a>0$, depending 
on the class of initial conditions considered. However,
by lowering $N$ we cross to an energy regime where the 
system  becomes $N$--depended and its corresponding time-scales  
extend exponentially like
$\tau \sim  \exp \varepsilon ^{-b(N)}$, $b(N)>0$.
Such a phenomenon seems to disappear in the thermodynamic limit
for two cases of random initial data with non-equipartitioned 
energy spectra.

We finally examine the information drawn from computing time fluctuations
of $J$ in cases in which the system is initially at energy equipartition,
with the initial conditions being distributed either far or close to the
Gibbs measure. In such cases, through $J(t)$ we can clearly compute a
speed of diffusion transversally to the Toda integral surfaces.
This speed is always small, implying that an underlying nearly-integrable
dynamics holds even when the initial conditions are close to
equipartition, and even close to the final equilibrium state.

As a final comment, we expect similar features, as those encountered
for the evolution of the first Toda integral, to appear also for the
other Toda integrals, a subject currently under investigation. In fact,
treating such integrals analogously as in Eq.(\ref{Jnm}) allows to
conjecture that, at low energies, this behavior will be reflected
satisfactorily by their harmonic approximations. 
A detailed investigation of this subject is proposed for
future study.

\section{Acknowledgments}
We are indebted to G. Benettin and A. Ponno for very fruitful discussions which significantly
improved the present work. 
H.C. was supported by the State Scholarship Foundation (IKY) 
operational Program: `Education and Lifelong Learning--Supporting Postdoctoral Researchers'
 2014-2020, and is co--financed by the European Union and Greek national funds.

\appendix

\section{Analytic estimates \label{analytic}}

The quadratic terms of (\ref{j1}) derive from the terms 
$T_1=\frac{ 1 }{2}a_n^2 (p_n^2 + p_n p_{n+1}+ p_{n+1}^2 )$ and
$T_2=\frac{a_n^2}{16 \alpha ^2}  ( a_{n+1}^2+a_n^2 +a_{n-1}^2)$. 
$T_1$ immediately yields the terms $\frac{ 1 }{2} (p_n^2 + p_n p_{n+1}+ p_{n+1}^2 )$
and $T_2$ yields the terms 
$\frac{1}{8} [(q_{n+2}-q_{n})^2 +4 (q_{n+1}-q_n)^2  + (q_{n+1}-q_{n-1})^2]$
after Taylor--expanding any of the terms $a_n^2a_{n+1}^2 = e^{ \alpha (q_{n+2}-q_n) }$ etc.

Now summing all quadratic terms together,
\begin{eqnarray}\label{qt}
QT= \frac{1}{N}\sum_n\frac{ 1 }{2} (p_n^2 + p_n p_{n+1}+ p_{n+1}^2 ) + \frac{1}{8} [(q_{n+2}-q_{n})^2 
+4 (q_{n+1}-q_n)^2  + (q_{n+1}-q_{n-1})^2]
\end{eqnarray}
their expression easily simplifies by matching terms 
like $\frac{ 1 }{2} \sum_n [p_n^2 + (q_{n+1}-q_n)^2] $
which can be replaced by $\sum _k E_k \simeq E$. 
Furthermore, it is $\sum_n (q_{n+2}-q_{n})^2 =\sum_n (q_{n+2}-q_{n+1}+q_{n+1}-q_{n})^2 
\approx 2\sum_n [\delta q_{n} \delta q_{n+1} + (q_{n+1}-q_{n})^2]$
and $\sum_n (q_{n+1}-q_{n-1})^2 \approx 2\sum_n [\delta q_{n} \delta q_{n+1} + (q_{n+1}-q_{n})^2]$
 which approximates the quadratic terms in (\ref{qt}) and gives the expression (4):
$$QT\approx  2\varepsilon  + \frac{ 1 }{2N} \sum_n ( p_n p_{n+1}+ \delta q_{n} \delta q_{n+1} )$$.
 
Under the Fourier transform the above expression takes the form (5). In particular,
it is $\sum_n p_n p_{n+1} = \sum_k (1-\frac{\omega _k^2}{2})P_k^2 = \sum_k 2 \cos (\frac{\pi k}{N})P_k^2$
and $\sum_n \delta q_{n} \delta q_{n+1} \approx 2 \sum_k \omega _k^2 \cos (\frac{\pi k}{N})Q_k^2 + 2 \sum_{l,m}c_{l,m} Q_lQ_m$,
where $c_{l,m}=\sin(\frac{l \pi}{N})\sin(\frac{m \pi}{N})$ and the off--diagonal sum $\sum_{l,m}c_{l,m} Q_lQ_m$ 
is approximately zero.
Therefore, we get that 
 $$QT\approx  2\varepsilon  + \frac{ 1 }{N} \sum_k \cos (\frac{\pi k}{N})E_k~~.$$
 

\nocite{*}


\end{document}